\documentclass[sigconf]{acmart}
\AtBeginDocument{%
  \providecommand\BibTeX{{%
    \normalfont B\kern-0.5em{\scshape i\kern-0.25em b}\kern-0.8em\TeX}}}

\copyrightyear{2024} 
\acmYear{2024} 
\setcopyright{rightsretained} 
\acmConference[CHI '24]{Proceedings of the CHI Conference on Human Factors in Computing Systems}{May 11--16, 2024}{Honolulu, HI, USA}
\acmBooktitle{Proceedings of the CHI Conference on Human Factors in Computing Systems (CHI '24), May 11--16, 2024, Honolulu, HI, USA}
\acmDOI{10.1145/3613904.3642195}
\acmISBN{979-8-4007-0330-0/24/05}

\usepackage{graphicx,subcaption} 

\begin{document}

\title[AR for First Responders]{Exploring the Design Space of Optical See-through AR Head-Mounted Displays to Support First Responders in the Field}


\author{Kexin Zhang}
\affiliation{
    \institution{University of Wisconsin-Madison}
    \city{Madison}
    \state{Wisconsin}
    \country{USA}
}
\email{kzhang284@wisc.edu}

\author{Brianna Cochran}
\affiliation{
    \institution{University of Wisconsin-Madison}
    \city{Madison}
    \state{Wisconsin}
    \country{USA}
}
\email{bcochran2@wisc.edu}

\author{Ruijia Chen}
\affiliation{
    \institution{University of Wisconsin-Madison}
    \city{Madison}
    \state{Wisconsin}
    \country{USA}
}
\email{ruijia.chen@wisc.edu}

\author{Lance Hartung}
\affiliation{
    \institution{University of Wisconsin-Madison}
    \city{Madison}
    \state{Wisconsin}
    \country{USA}
}
\email{lhartung@wisc.edu}

\author{Bryce Sprecher}
\affiliation{
    \institution{University of Wisconsin-Madison}
    \city{Madison}
    \state{Wisconsin}
    \country{USA}
}
\email{bjsprecher@wisc.edu}

\author{Ross Tredinnick}
\affiliation{
    \institution{University of Wisconsin-Madison}
    \city{Madison}
    \state{Wisconsin}
    \country{USA}
}
\email{rdtredinnick@wisc.edu}

\author{Kevin Ponto}
\affiliation{
    \institution{University of Wisconsin-Madison}
    \city{Madison}
    \state{Wisconsin}
    \country{USA}
}
\email{kbponto@wisc.edu}

\author{Suman Banerjee}
\affiliation{
    \institution{University of Wisconsin-Madison}
    \city{Madison}
    \state{Wisconsin}
    \country{USA}
}
\email{suman@cs.wisc.edu}

\author{Yuhang Zhao}
\affiliation{
    \institution{University of Wisconsin-Madison}
    \city{Madison}
    \state{Wisconsin}
    \country{USA}
}
\email{yuhang.zhao@cs.wisc.edu}

\renewcommand{\shortauthors}{Zhang et al.}
\newcommand{\change}[1]{{\color{black}#1}}
\newcommand{\yuhang}[1]{{\small\textcolor{red}{\bf [#1]}}}
\newcommand{\kexin}[1]{{\color{magenta} \textbf{(Kexin: #1)}}}

\begin{abstract}
First responders (FRs) navigate hazardous, unfamiliar environments in the field (e.g., mass-casualty incidents), making life-changing decisions in a split second. 
AR head-mounted displays (HMDs) have shown promise in supporting them due to its capability of recognizing and augmenting the challenging environments in a hands-free manner. However, the design space have not been thoroughly explored by involving various FRs who serve different roles (e.g., firefighters, law enforcement) but collaborate closely in the field. 
We interviewed 26 first responders in the field who experienced a state-of-the-art optical-see-through AR HMD, as well as its interaction techniques and four types of AR cues (i.e., overview cues, directional cues, highlighting cues, and labeling cues), soliciting their first-hand experiences, design ideas, and concerns. Our study revealed both generic and role-specific preferences and needs for AR hardware, interactions, and feedback, as well as identifying desired AR designs tailored to urgent, risky scenarios (e.g., affordance augmentation to facilitate fast and safe action). While acknowledging the value of AR HMDs, concerns were also raised around trust, privacy, and proper integration with other equipment. Finally, we derived comprehensive and actionable design guidelines to inform future AR systems for in-field FRs.
\end{abstract}






\maketitle


%
\section{Introduction}

First responders (FRs) refer to people who are employed or volunteer to respond to fire, medical, hazardous material, or other emergencies~\cite{legalinformation}. FRs include different disciplines, such as firefighters (FF), law enforcement (LE), emergency medical services (EMS), and communications center \& 9-1-1 services~\cite{legalinformation}, 
who engage in a wide range of tasks, ranging from low-risk routine activities like traffic stops \cite{grandi_2021_Design} and responding 911 calls \cite{legalinformation}, to high-stakes scenarios, such as dealing with active shooter \cite{callaway2024active}, locating and rescuing people from burning buildings \cite{phillips2016structure}, and triaging in mass-casualty events \cite{alpert2023emsmass}. \change{Among all FR disciplines, three disciplines---FF, LE, and EMS---commonly work \textit{in the field}, directly confronting emergencies and performing various on-site duties at incident locations \cite{dawkins2021futuristic}. Tasks in the field are particularly challenging and dangerous with numerous unpredictable and rapidly evolving hazards, demanding extra physical and mental efforts from the FRs to make life-or-death decisions within seconds \cite{dawkins2018public, killeen2006}.} 
Unfortunately, technologies to support \change{in-field emergency response} are very limited. For example, despite the advance of communication technologies in daily life, FRs are mostly using land mobile radios for communication \cite{dawkins2018public}, and not all FRs even have access to these radios \cite{dawkins2021futuristic}. 


Augmented reality (AR) \change{head-mounted displays (HMDs) (e.g., HoloLens 2)} have been recognized as a promising platform to support FRs by intelligently recognizing the real-world environment and rendering in-situ augmentations, facilitating performance and safety in the field \cite{yohan2000bars}.  
\change{Compared to mobile AR, head-mounted AR presents a head-up overlay directly onto the real world, }
avoiding unnecessary attention switch between the physical environment and any handheld devices; the head-mounted form factor also frees FRs' hands for more important primary tasks in urgent scenarios \cite{datcu2016handheld}. 
\change{Despite the potential, AR HMDs, especially the optical see-through HMDs, can bring new issues and risks, such as additional burden on FRs' head, limited field of view, distractions, and visibility issues of the virtual elements against the real-world background \cite{cheng2021design, vankrevelen2010survey, cakmakci2006headworn}
, which may significantly affect FRs' experiences and safety in the field.
However, while mobile AR for FRs has received certain attention in the literature \cite{Sebillo2016, Nunes_2018_ARapp, AugView, datcu2016handheld, siu2013sidebars}, limited research has investigated the unique opportunities and challenges posed by head-mounted AR for FRs in the field \cite{kapalo2018sizing, kapalo2018sizing, wilson2005design}, and even less work has situated FRs in an AR context to understand their first-hand experiences \cite{chan2016erwear}.}



To fill this gap, we adopt a user-centered approach to deeply understand FRs' needs and concerns with using head-mounted AR devices in the field. We comprehensively explore the design space from three perspectives, including hardware, interaction techniques (e.g., hand gestures, voice commands, gaze input), and AR feedback. As emergency response usually requires the collaboration of multiple FR roles in the field \cite{kristiansen2018whenitmatters, ARTTS2022}, 
we also involve different FR disciplines (i.e., FF, LE, EMS) in our research to explore their different needs. We seek to answer below research questions:

\change{\textbf{RQ1}: For hardware, what are the benefits and drawbacks of the AR HMD form factor for FRs in the field? }

\change{\textbf{RQ2}: For interaction techniques, what are the benefits and drawbacks of different interaction techniques for FRs? What interaction techniques are preferred in different emergency response scenarios by different FR roles? }

\change{\textbf{RQ3}: For AR feedback, what are the key information to recognize and augment in the field? What forms of AR augmentations are preferred in different emergency response scenarios by different FR roles? }

To achieve this goal, we provide FRs \textit{first-hand} experiences with \change{a state-of-the-art AR HMD---HoloLens 2---}and conduct semi-structured interviews with 26 participants that cover three in-field FR roles, including FF, LE, and EMS. To better situate in AR context, \change{all participants wear HoloLens 2 and experience its interaction techniques. They also explore} four types \change{of widely used} AR cues as the design probes (18 participants viewed the cues via HoloLens 2 and eight watched videos that demonstrated the cues). The AR cues included: (1) \textit{overview cues}: 2D and 3D maps; (2) \textit{labeling cues}: world-anchored icons and text;  (3) \textit{directional cues}: arrow and path; and (4) \textit{highlighting cues}: contour and shaded overlay on real-world objects. Participants discuss their preferences and concerns, as well as brainstorming suitable scenarios and AR ideas based on the design probes. 


\change{Via a deep discussion with diverse FRs, we identified not only generic implications applied to broad emergency response, but also scenario-specific or discipline-specific insights tailored for individual FR roles. 
For example, although most FRs found AR HMD support stable wear, FFs needed them to be extra tight for tasks involved large body movements. While all FRs favored HoloLens 2' speech commands for hands-free operation and minimal distraction, other interaction modalities, such as gaze and vibration were useful for compensating specific challenging scenarios, such as noisy environment or extreme dangerous conditions. Moreover, we identified FRs' different design and usage preferences for AR feedback. 
For instance, while FFs desired visually salient directional cues for back-tracing in smoke-filled environment, more subtle directional cues should be adopted to support emergency vehicle driving, avoid blocking real-world elements. 

In summary, our research makes three contributions. First, we conducted an user-centered, in-depth investigation to comprehensively explore the design space of AR HMDs in the field. Second, we provided FRs with first-hand experience and contextualized them in AR context, enabling deep and thorough understandings of AR HMD technologies. Third, 
we developed actionable design guidelines from different perspectives to guide both general and role-specific use of AR HMDs in emergency response.}

\section{Related Work}


\subsection{Challenges Faced by First Responders}
A myriad of research has studied FRs' challenges and needs. Maintaining situational awareness in dynamically changing environments was one significant challenge encountered by 
multiple types of FRs \cite{jiang2004siren, Kyng_2006_challenges, grandi_2021_Design, killeen2006, denef2009letting, Lorincz_2004_sensor}. For example,  
Grandi et al. \cite{grandi_2021_Design} investigated the traffic stop, a routine task where the police officer approached the driver for identity verification, and revealed the risks of unknowingly engaging with individuals involved in criminal activities. Killeen et al. \cite{killeen2006} highlighted paramedics' challenges of tracking victims' latest information to determine medical resource needs.
Compared to other FR roles, firefighters often operated in low-visibility environments due to heavy smoke and steam-fogged lenses on firefighting gears \cite{grandi_2021_Design, jiang2004siren, yang2009}, leading to additional challenges in situational awareness and navigation. Through a 4-month field study and interviews with 14 firefighters, Jiang et al. \cite{jiang2004siren} found that the firefighters had to explore the scene blindly using non-traditional locomotion style (e.g., touching and counting walls) and it was difficult to gather accurate incident information due to the invisible environment and dynamic nature of fire \cite{grandi_2021_Design, jiang2004siren}.

Communication was another significant challenge in emergency response \cite{toups2007implicit, jiang2004siren, scholz2013concept, manoj_2007_comm_challenges, dawkins2021futuristic}. Toups et al. \cite{toups2007implicit} identified two communication methods in the field: face-to-face interactions and shared radio communication. However, face-to-face communications often had limited availability due to background noises on incident scenes \cite{jiang2004siren}, and the radio technology was not always reliable with multiple failures, such as signal breakdowns, wireless interference, and dead batteries \cite{scholz2013concept}. The limited bandwidth and insufficient number of radio channels added extra burden to FRs \cite{Lorincz_2004_sensor}. Moreover, the verbal communication cannot effectively convey visual information (e.g., maps), preventing the FRs from establishing a shared understanding of the incident scene \cite{Kyng_2006_challenges, MONARES_2011_mobile}. 

\subsection{First Responders' Experiences with Technologies}
While little technology has been used by FRs in the field, researchers have started exploring the potential of specific technologies for first responder \cite{amon2005thermal,alon2021drones,khan2019exploratory}. For example, Amon et al. hosted a workshop with 38 attendees that were FRs and thermal imaging manufacturers to discuss strategies, technologies, procedures, and best practices to improve firefighter's safety and fire protection, as well as the effectiveness of thermal imaging in the fireground \cite{amon2005thermal}. Alon et al. interviewed three firefighters to understand the usage and barriers of drones. They also conducted a focus group of nine firefighters and a gesture elicitation study to investigate collocated human-drone interactions to support FF~\cite{alon2021drones}. 

There has been limited research that studied FRs' needs for AR devices. Specifically, Kapalo et al. interviewed 35 firefighters and 3 emergency management personnel to understand their challenges and identify how AR could be used to support them in the field \cite{kapalo2018sizing}.
Wilson et al. interviewed 50 firefighters about their needs for head-mounted displays and revealed the drawbacks of the devices, such as blocking lower vision, stop functioning in cold weather \cite{wilson2005design}. However, these works only asked participants to imagine the experiences of using AR glasses and answer hypothetical questions.    
Chan et al.'s work was the only exception. They interviewed 21 FRs and gave them a Recon Jet eyewear to discuss their experiences and feedback \cite{chan2016erwear}. However, unlike the state-of-the-art AR glasses that can register 3D virtual elements in the physical environment, the eyewear in this study was consisted of a pair of sunglasses with a small display attached to the front of the right eye, which functioned as an external phone screen instead of augmenting the physical world directly. The researchers found that such devices required repeated attention switches between the display and the physical world. No research has explored FRs' first-hand experiences with the modern AR glasses. 



\subsection{AR Technology to Support First Responders}
Extensive efforts have been made in creating technologies to support FRs in various tasks, including communication \cite{farnham2006observation, aedo2010end, prasanna2011evaluation, camp2000supporting, streefkerk2006designing}, 
navigation \cite{klann2007lifenet,klann2011experience, steingart2005augmented, liu2010automatic, ramirez2010designing}, and information gathering and sharing \cite{jiang2004ubiquitous, betz2014emergencymessenger, kasnesis2022deep, streefkerk2008field, purohit2011sensorfly, beata2018real}. 
For example, Bergstrand et al. ~\cite{bergstrand2009using} developed a live response application for FRs that incorporated map services, live video support, and awareness features. 
Kasnesis et al. proposed a back-mounted wearable device, which can recognize dog activity, barks, thus providing the dog handler and the dog's location through a mobile application~\cite{kasnesis2022deep}. Moreover, Jiang et al. \cite{jiang2004siren} presented Siren, a peer-to-peer context-aware computing architecture that gathered contextual data to support spontaneous interactions between firefighters. 

With the advancement of AR technology, research efforts have been made to design and develop AR systems to support FRs \change{in the field}, mostly focusing on hand-held AR on mobile devices \cite{chen_2020_equipment, Sebillo2016, sebillo2015use, Nunes_2018_ARapp, AugView, datcu2016handheld, siu2013sidebars}. 
For example, Chen et al. \cite{chen_2020_equipment} designed a mobile AR framework to facilitate fire safety equipment inspection. 
Nunes et al. \cite{Nunes_2018_ARapp} presented THEMIS-AR, a mobile AR application that augmented FRs' scene perception by overlaying context information and guidance (e.g., symbols, pictures, text) to the video stream captured through a smartphone camera. Moreover, Campos et al. ~\cite{campos2019mobile} created an AR mobile application to assist FRs in emergencies by displaying points of interests and a mini-map. However, such handheld systems occupy users' hands and require them to constantly switch focus between the screen and the physical environment \cite{datcu2016handheld}. 




As opposed to mobile AR, \change{head-mounted AR} frees FRs' hands for more important primary tasks and avoids unnecessary attention switch between the real-world environment and any portable devices \cite{datcu2016handheld}. Only a small number of research has proposed or developed head-mounted AR applications for FRs \cite{wani2013augmented, brunetti2015smart, sainidis2021single, kamat2007evaluation}. 
For example, Saindis et al.~\cite{sainidis2021single} presented a system that enabled FRs to intuitively control an unmanned aerial vehicle (UAV) through single-hand gestures and visualized the UAV’s camera feed on HoloLens 2. However, these works only focused on the technical framework and development, without involving FRs in the system design and evaluation. \change{One exception is a recent research by Nelson et al. \cite{ARTTS2022, Nelson_2022_explore}, who designed a set of head-worn AR triage tools to assist FRs in massive casualty incidents via user-centered approaches, involving FRs in the iterative prototyping and evaluation process. However, this work specifically focused on the triaging scenarios and did not consider other in-the-field tasks.} 
\change{More in-depth and thorough} research is needed to understand different FRs' first-hand experiences \change{and concerns} with state-of-the-art \change{head-mounted AR technologies}, thus designing effective and safe AR systems to support \change{their operation and collaboration} in the field.

\section{Method}
We seek to investigate FRs' needs and preferences for \change{head-mounted AR} technologies and derive design guidelines \change{from three aspects---hardware form factors, interactions, and design of AR cues}. To situate participants in the AR context, \change{we used the Microsoft HoloLens 2, its commonly used interaction techniques (i.e., in-the-air gestures, voice command, gaze input), and four type of AR cues (i.e., overview, directional, labeling, and highlighting cues) as \textit{design probes}} in an in-depth semi-structured interview study. 
This study was approved by the Institutional Review Board (IRB) at the University of Wisconsin-Madison. 

\subsection{Participants}
We recruited 26 FRs (5 female, 21 male) with experiences ranging from 4 months to 32 years. \change{As handling in-the-field tasks usually requires the involvement and collaboration of multiple FR roles \cite{ARTTS2022, kristiansen2018whenitmatters, manoj2007emergencyresponse}, 
we recruited FRs in different roles to develop a comprehensive understanding of the needs of different types of FRs.} Our participants thus covered three FR disciplines\change{---EMS, FF, LE---who  commonly operate in the field \cite{dawkins2021futuristic}
}, including seven FF, six LE officers, six EMS (all of them were paramedics), and seven participants who were both FF and EMS. Participants had various duties, such as 911 calls responding, firefighting, urban search and rescue, driving ambulances, emergency patient care, hostage rescues, and active shooters. \change{Only six participants had prior experiences with AR, two (P1, P2) of them tried head-mounted AR glasses (e.g., Google Glasses, HoloLens 2) a few times at FR conferences, two (P7, P12) had a one-time experience with AR training applications about vascular structure visualization, P12 had gaming experiences with AR, and P17 tried mobile AR once.} Table \ref{tab:participants} shows participants' demographic information. 

\begin{table*}[h!]
    \footnotesize 
    \centering
    \caption{Participants' Demographics, including ID, age, gender, yeas of experience, and occupation. \change{* denotes participants who did not experience the AR cues on HoloLens 2 but watched the demo video of the AR cues; however, they worn HoloLens 2 and experienced the interaction techniques via Tip Application as other participants did.}} 
    \begin{tabular}{p{0.3cm}p{1cm}p{1.5cm}p{3cm}}
    \toprule
        \textbf{ID} & \textbf{Age/ Gender} & \textbf{Yeas of Experience} & \textbf{Occupation}\\
        \toprule
        \textbf{P1} & 45/M & 23 years & Firefighter; Paramedic \\
        \midrule
        
        \textbf{P2} & 43/M & 12 years & Paramedic \\
        \midrule
        
        \textbf{P3} & 37/M & 12 years & Firefighter; Paramedic \\
        \midrule
        
        \textbf{P4} & 40/M & 15 years & Firefighter; Paramedic \\
        \midrule
        
        \textbf{P5} & 31/M & 13 years & Firefighter; Paramedic \\
        \midrule
        
        \textbf{P6} & 46/M & 15 years & Firefighter; Paramedic \\
        \midrule
        
        \textbf{P7} & 20/F & 4 months & Paramedic \\
        \midrule
        
        \textbf{P8} & 28/F & 1 year & Firefighter; Paramedic \\
        \midrule
        
        \textbf{P9} & 45/M & 8 years & Paramedic \\
        \midrule
        
        \textbf{P10} & 57/M & 38 years & Lieutenant at Fire Department \\
        \midrule
        
        \textbf{P11} & 52/M & 32 years & Fire Apparatus Engineer \\
        \midrule
        
        \textbf{P12} & 36/M & 18 years & Paramedic \\
        \midrule
        
        \textbf{P13} & 35/M & 15 years & Firefighter; Paramedic \\
        \bottomrule
    \end{tabular}
    \quad
    \begin{tabular}{p{0.3cm}p{1cm}p{1.5cm}p{3cm}}
    \toprule
        \textbf{ID} & \textbf{Age/ Gender} & \textbf{Yeas of Experience} & \textbf{Occupation}\\
        \toprule
        \textbf{P14} & 50/M & 7 years & Firefighter \\
        \midrule
        
        \textbf{P15} & 43/F & 8 years & Paramedic \\
        \midrule
        
        \textbf{P16} & 52/F & 24 years & Lieutenant at Fire Department \\
        \midrule
        
        \textbf{P17} & 36/M & 10 years & Paramedic \\
        \midrule
        
        \textbf{P18} & 29/F & 8 years & Police Officer \\
        \midrule
        
        \textbf{P19*} & M & 4 years & Firefighter \\
        \midrule
        
        \textbf{P20*} & M & 4 years & Firefighter \\
        \midrule
        
        \textbf{P21*} & M & N/A & Firefighter \\
        \midrule
        
        \textbf{P22*} & M & N/A & Law Enforcement\\
        \midrule
        
        \textbf{P23*} & M & 11 years & Law Enforcement\\
        \midrule
        
        \textbf{P24*} & M & N/A & Law Enforcement\\
        \midrule
        
        \textbf{P25*} & M & N/A & Law Enforcement\\
        \midrule
        
        \textbf{P26*} & M & N/A & Law Enforcement\\
        \bottomrule
    \end{tabular}
    \label{tab:participants}
\end{table*}

We leveraged multiple channels for recruitment, including mailing to local public safety organizations and technical colleges, visiting local fire departments and police departments, posting recruitment message on online FR communities on mainstream social media platforms (e.g., Facebook, LinkedIn), and referrals from recruited participants. Interested participants would fill out a survey with screening questions, in which we asked about participant's age, current occupation, years of experiences as a FR, and their availability for a in-person study. Eligible participants must: (1) be over 18 years old and (2) have experiences working as a FR. \change{All participants served in the United States. To achieve a good amount of FRs that cover different roles, our recruitment lasted for two years, from November 2021 to July 2023.} 

\subsection{Apparatus: Design Probes in AR}
\change{To provide participants first-hand experiences with AR HMDs, we presented them a pair of HoloLens 2 and asked them to experience various AR cues and interaction techniques on it. We introduce the three types of design probes used in our study, including the hardware, the interaction techniques, and AR cues below.} 

\subsubsection{\change{AR Device: HoloLens 2.}} We used Microsoft HoloLens 2 as the study platform since it is one of the \change{most} state-of-the-art \change{optical see-through (OST)} AR HMD in the market. \change{In contrast to video see-through (VST) HMDs that replace users' natural vision with video streaming of the surroundings, the OST HMDs directly register virtual elements in the real world space and do not block users' original vision \cite{zhou2008trends}. As such, the OST HMDs can diminish disorientation and motion sickness \cite{zhou2008trends}, have higher error tolerance (e.g., not blocking users' vision when running out of power), and are more socially acceptable \cite{mp2020functionality, zhao2017understanding}, thus being more suitable for FRs in urgent and challenging MCIs. In fact, HoloLens 2 has already provided a hard-hat integration for personnel in dirty, loud, and safety-controlled work site environments (i.e., \textit{Trimble XR10}) \cite{TrimbleXR10}.} Besides the form factors, HoloLens 2 has powerful computation and \change{sensing capabilities for real world augmentations} \cite{MicrosoftHololens2}. It has multiple input sensors to perceive the real world and recognize the user's behaviors, including an RGB camera, a depth sensor, four visible light cameras and an inertial measurement unit (IMU) for head tracking, two IR cameras for eye tracking, and a microphone array for voice command. For output, it has a pair of see-through holographic lenses to render 3D virtual elements and speakers to generate spatial audio \cite{HololensSpecs}. We introduce the two HoloLens 2 applications used in our study. 

\subsubsection{\change{Interaction Techniques.}} \label{sec:tip}
\begin{figure*}
     \centering
     \includegraphics[width=\linewidth]{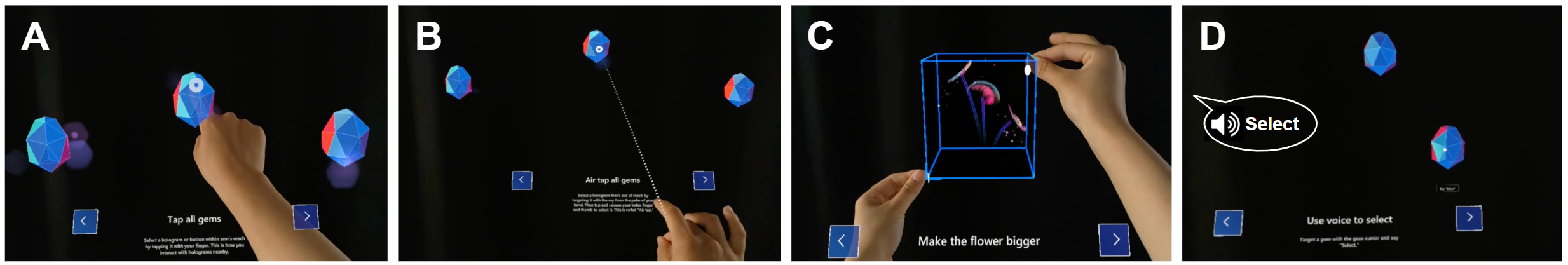}
     \caption{(A) Tap to select holograms within reach; (B) Air Tap to select holograms out of reach; (C) Pinch to resize holograms; (D) Select holograms using eye cursor and a voice command.}
     \Description{The image displays four panels, A, B, C, and D, each showing a hand interacting with a graphical user interface that features floating, multicolored gem-like objects against a dark background.
     Image A: A hand is tapping a gem. Text reads "Tap all gems" with additional instructions indicating to select a hexagon gem button and to reach targets by tapping or with hand gestures nearby.
     Image B: A finger is pointing, with a dashed line from the fingertip to a gem, indicating a tap action. Text reads "Air tap all gems" with similar additional instructions as in panel A.
     Image C: Two hands are pinching and stretching apart on a digital box that encompasses a flower inside it, with text that reads "Make the flower bigger".
     Image D: A hand with an extended index finger is positioned near a gem. An icon with a voice command symbol and text "Select" is displayed. Additional text reads "Use voice to select" with instructions to target a gem with the same color and say "Select".}
     \label{fig:Tips}
\end{figure*}

We used the system embedded ``Tip'' application ~\cite{HoloLensInteraction} on HoloLens 2 as a tutorial to demonstrate the multi-modal interactions afforded by HoloLens 2. The application introduced various interaction techniques, including in-the-air gestures (e.g., air tap to select, move, resize, rotate, Fig. \ref{fig:Tips}A-C), voice commands, and gaze cursor via eye tracking (Fig. \ref{fig:Tips}D). In the app, a user was asked to interact with various holograms using these interaction techniques, for example, looking at a hologram and saying ``select'' to select the hologram (Fig. \ref{fig:Tips}D).


\begin{figure*}
     \centering
     \includegraphics[width=\linewidth]{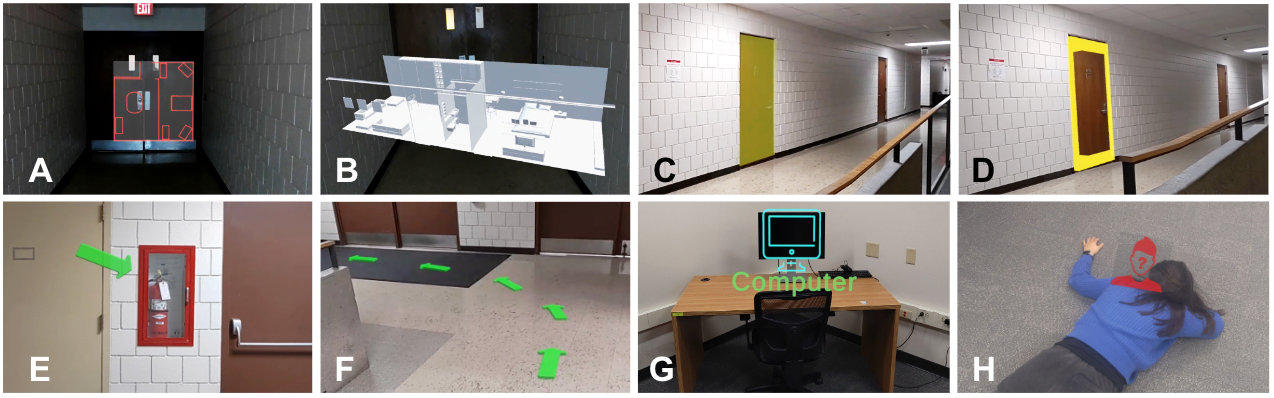}
     \caption{\textbf{Overview cues:} (A) Head-attached map; (B) 3D Floor plan. \textbf{Highlighting cues:} (C) Door with shade; (D) Door with Contour. \textbf{Directional cues:} (E) Single arrow pointing a POI; (F) An array of arrows indicating a path. \textbf{Labeling cues:} (G) Icon and text labeling an object; (H) Icons labeling a victim.}
     \Description{The image is a compilation of eight panels, labeled A to H, showing various indoor scenarios with four types of AR cues providing instructions or information.
     A: A dark hallway with an EXIT sign above the door at the end. Red outlines representing the map of this space are superimposed on elements of the door.
     B: A schematic overlay on a perspective of a room, showing a white, digital model of an office with desks and chairs. This could be a planning tool or a virtual representation of space.
     C: A hallway with a door highlighted in yellow shades. 
     D: The same hallway as in C, but the door is now highlighted in yellow, thick outlines.
     E: A close-up of a fire extinguisher inside a red box on a wall, with a green arrow pointing towards it.
     F: A floor with green arrows leading towards a destination.
     G: An office desk with a computer, where the computer screen has a green overlay with an icon and the word "Computer".
     H: A person is lying face down on the floor, seemingly unconscious, with a red warning icon and a question mark above them, indicating an emergency situation that requires attention.}
     \label{fig:probes}
 \end{figure*}

\subsubsection{Four Types of AR Cues.} \label{sec:cues}
\change{We also designed and implemented four types of AR cues for participants to explore and discuss. To ensure our AR cues cover a comprehensive set of design forms and use cases, we conducted a systematic literature review on existing AR research for FRs. Our literature search included both general query on Google Scholar and more targeted search on representative HCI and AR/VR venues (e.g., ACM CHI, UIST, CSCW, UbiComp, IEEE VR, ISMAR). The keywords we used included ``first responders,'' ``emergency response,'' ``emergency management,'' or ``disaster response,'' combined with ``challenges,'' ``needs,'' ``in-the-field,'' ``augmented reality (or AR),'' ``AR designs,'' ``AR prototypes,'' ``AR applications,'' or ``AR interfaces.'' We further filtered the searched results by manually examining the full text of each paper and identifying irrelevant and repetitive papers. 

As a result, we collected a final list of 27 papers that focused on AR technologies for FRs, and identified four types of AR cues, including: 1) maps for space visualization and overview \cite{campos2019mobile, Weichelt_2018_Augmented, Vassell_2016_intelligent, Peretti_augmented_2022, chan2016erwear, fromm2023social, grandi_2021_Design, nunes2017augmented, koutitas2020smart, Sharma_2020_situational}; 2) icons and text to label point of interests and provide detailed information \cite{Nunes_2018_ARapp, grandi_2021_Design, fromm2023social, Weichelt_2018_Augmented, campos2019mobile, agrawal2021rescuear, Mehta_Human_2022, Vassell_2016_intelligent, koutitas2020smart, Sebillo2016, void_Hu_2022, Peretti_augmented_2022}; 3) arrows and paths for navigation \cite{Sharma_2020_situational, nunes2017augmented, grandi_2021_Design, fromm2019potential}; and 4) outline and shaded overlay for object highlighting and enhancement \cite{Bhattarai_2022_Deep, grandi_2021_Design}.} 


\change{Based on the literature review, we designed and implemented four types of AR cues accordingly as design probes}: (1) \textbf{overview cues} that visualize the overview of a site via a head-attached 2D map (Fig. \ref{fig:probes}A) \change{and} a 3D floor plan (Fig. \ref{fig:probes}B) \cite{grandi_2021_Design, Sharma_2020_situational, lalone_2019_vision, campos2019mobile, Weichelt_2018_Augmented, Vassell_2016_intelligent}; a user can use in-the-air gestures to interact with the map, such as moving, resizing, and rotating; (2) \textbf{directional cues} that indicate a target or a direction via a single arrow (Figure \ref{fig:probes}E) \change{and} an array of arrows connecting into a path (Figure \ref{fig:probes}F) \cite{Sharma_2020_situational, nunes2017augmented, fromm2019potential}; (3) \textbf{highlighting cues} that emphasize an object of interest by adding a colored shaded overlay (Fig. \ref{fig:probes}C) or a contour \change{around the object} (Fig. \ref{fig:probes}D) \cite{Bhattarai_2022_Deep}; and (4) \textbf{labeling cues} that labels objects or human of interest with icons and text (Fig. \ref{fig:probes}G-H) \cite{Nunes_2018_ARapp, grandi_2021_Design, fromm2023social, Weichelt_2018_Augmented, campos2019mobile, agrawal2021rescuear}. We implemented the AR cues on HoloLens 2 using the Mixed Reality Feature Tool \cite{MRFT} in Unity 2020.3.33. 


\subsection{Procedure}
\change{We conducted a semi-structured interview study with design probes with each participant. Each study consisted of a single session} that lasted two hours. 
We first asked about participants' demographics (i.e., gender), current occupation, and years of experiences as a FR. Participants then introduced their work duties, the challenges they face, technologies they used in the field, and their prior experiences with \change{AR technologies}. \change{We then asked them to explore and assess HoloLens 2's hardware components, the interaction techniques it supported, and the four types of AR cues through the following three phases. The detailed study protocol could be found in \ref{protocol}.}

\subsubsection{\change{Hardware Assessment.}} We first presented HoloLens 2 to the participants and introduced its hardware features, including both the input sensors and the output components, to help them understand the device capability. Participants then put on the HMD and experienced 
\change{its physical form factor (e.g., weight, comfort). Participants can freely walk around wearing the HoloLens 2; no AR element were presented at this stage}. We asked participants about their wearing experiences, such as whether the HMD felt heavy to wear, any restrictions to body movement, influences to \change{their natural} field of view, safety regarding emergency response standards, and concerns about wearing it in the field. 

\subsubsection{\change{Interaction Technique Exploration.}} Participants then experienced three types of interaction techniques on HoloLens 2 by using the Tip application, including in-the-air gestures, voice commands, and gaze-based interaction (Section \ref{sec:tip}). 
The exploration lasted about 10 minutes. We then asked about participants' experiences with each interaction technique, including the usability, learnability, their willingness to use such an interaction in the field, and suggestions to improve the technique. 

\subsubsection{\change{AR Cue Exploration.}} After fully understanding HoloLens 2' capability and interactions, we conducted a brainstorming session with the four types of AR cues as design probes (Section \ref{sec:cues}). Participants wore the HoloLens 2 and experienced each AR cue. 
While most participants experienced the cues in HoloLens 2, eight participants (P19-26) watched the demo video of the cues since we visited their sites in-person and could not easily set up the spatial mapping of HoloLens 2 for the world-anchored cues. \change{Although not experiencing the AR cues on HoloLens 2, these participants put on the HoloLens 2 and experienced the interaction techniques via the Tip application on HoloLens 2 like other participants. As such, they still had a reasonable amount of first-hand experience with HoloLens 2 and its capabilities to provide valuable feedback. We did not observe different trends from these eight participants compared to others.} 

For each cue, we asked participants to brainstorm about whether this type of cue would be helpful, what tasks or information can be augmented by such a design, their preferences on different design aspects (e.g., placement, color) of the cue, how they want to improve the design, and what input techniques they want to use to interact with the cue. 
After discussing all the cues, we further prompted the participants to think about other possible AR cues they desire, including other feedback modalities (e.g., audio and haptic). 






We ended the study by asking participants' general feedback with the AR \change{HMD}, including its potential to be used in the field, their willingness to use it in the future, and their concerns with the technology.

\subsection{Data Analysis}
We video-recorded all studies and transcribed interviews using an automatic transcription service. \change{One researcher on the team reviewed the interview videos and the transcripts manually to correct errors caused by automatic transcription. We analyzed the transcripts using thematic analysis \cite{clarke2015thematic, braun2006using, fereday2006demonstratingrigor}.} 
\change{First, two researchers open coded} three identical samples (more than 10\% of the data) 
independently at the sentence level and developed an initial codebook by discussing and reconciling their codes to resolve any differences. \change{We assessed the intercoder reliability using Cohen's Kappa, resulting in a Kappa of 0.86, which indicated high agreement between two researchers.}
Next, the two researchers divided the rest of the transcripts and coded them independently \change{based on the initial codebook}. During this process, they regularly checked each other’s codes and discussed as needed to ensure consistency. New codes were added to the codebook upon agreement between the researchers. \change{Meanwhile, a third researcher oversaw all coding activities to ensure a high-level agreement. The final codebook consisted of 244 codes. The data saturated for all three FR types. 

We then derived themes from the codes by combining the inductive and deductive approaches \cite{braun2006using}. Our research has a specific goal to understand how head-mounted AR could support the needs of different FRs in the field from multiple aspects, including hardware assessment for practical in-the-field use, experiences with interaction techniques, and AR feedback on essential information to be augmented in the field. Our high-level theme generation thus focused on these particular aspects following the deductive approach. Under each theme, we then adopted the inductive approach and generated sub-themes by grouping relevant codes using axial coding and affinity diagram.  
After the initial themes and sub-themes were identified, researchers cross-referenced the original data, the codebook, and the themes, to make final adjustments, making sure that all codes fell in the correct themes. Our analysis resulted in seven themes with 24 sub-themes, 
and the details could be found in theme table in Appendix (Table \ref{tab:themes}).}


\section{Findings}









\subsection{Hardware Form Factor and Integration} \label{hardware}
Via the first-hand experiences with HoloLens 2, 15 out of 26 participants felt comfortable wearing HoloLens 2, reporting that the weight was barely noticeable for FRs who were used to heavy helmet (P7, P17). \change{However, one female participant (P15) found it too heavy for long wear and reported experiencing fatigue and a headache after wearing the device for two hours. Nevertheless,} the headband with adjustment wheel on HoloLens 2 enabled a good fit and stable wear without restricting FRs' body movement in urgent tasks (P1, P2, P3, P5, P14). 
\change{Besides the wearing comfort, fourteen participants 
appreciated the head-worn form factor of the HMD as opposed to handheld devices since they did not need to constantly alternate gaze and switch attention between a handheld screen and their physical surroundings---a common issue with their commonly used mobile devices, such as the thermal camera used by FFs to see objects in dark environment (P3, P15).}

However, the current hardware design also pose issues to FRs, especially in the hazardous and physically demanding environments. Participants \change{identified three issues with the HMD:}

\textbf{\textit{\change{Field of View.}}} 
\change{Ten participants reported issues caused by the limited field of view. The lower edge of the HMD blocked FRs's lower peripheral vision and caused bifocal view that impaired their judgement (P1, P10, P18). 
The small field also reduced the discoverability and visibility of AR cues anchored in the physical space (P6, P10, P11, P12, P15). 
As P15 described: \textit{``I have to have my head in the right spot to see the [AR] cues. If I look down too far, I can't see them; if I look up [too far], they're gone again.''} }

\change{Besides the visual experience issues, participants found the semi-enclosed HMDs not fulfill the safety standard for FRs.} Since FRs frequently encountered occupational exposure, such as high heat, smoke, and toxic chemicals for firefighters (P1) and blood splatter and vomit for paramedics (P1, P12, P16), 
participants (P1, P3, P16) emphasized the importance of \textit{``OSHA-approved''} hardware design, following the safety standards in Occupational Safety and Health Administration\footnote{OSHA. \url{https://www.osha.gov/sites/default/files/publications/osha3021.pdf}} that side shields are needed to protect eyes. 
As a result, nine participants suggested extending the AR display to be wraparound, providing full coverage around the eye area for both protection and wider field of view.


\textbf{\textit{\change{Stability of Device.}}} 
\change{Different FRs roles expressed different standards for the wearing stability of the HMD. All EMS and LE participants felt the AR HMDs tight and stable enough for common FR tasks such as foot patrol (P18) and driving emergency vehicles (P15). However, multiple firefighters (P4, P15, P17) were concerned that the HMD may fall off and needed extra stability, since they often worked in extreme situations involving large-scale body movements, such as crawling, leaning, and bending in confined space rescue. }
P4 and P17 thus suggested attaching a chin strap to the glasses to further stabilize the device. As P4 mentioned: \textit{``There would definitely need to be some sort of a strap component [attached to AR glasses]. Because I could see them falling off pretty easily [when] getting engaged in confined space rescue.''} 

\textbf{\textit{\change{Integration to Existing Equipment.}}} 
As opposed to solely wearing the AR HMD, ten participants preferred integrating AR capability into their current equipment. Four participants (P3, P5, P6, P15) pointed out that the \change{head-mounted AR} conflicted with the firefighter's helmet and face shields of Self-Contained Breathing Apparatus (SCBA). In addition, \textbf{wearing AR glasses would change the FRs' traditional appearance, which can cause social acceptance and trust issues.} Five participants (P1, P2, P8, P12, P13) concerned that the AR headset would confuse the community they served. As P8 reflected: \textit{``People want to be able to see your eyes. [Wearing this] could freak [people] out, because I'm not look like a normal person.''} Moreover, participants noted that the profession of emergency response was \textit{``very rooted in tradition''} (P5) and needed long-term trust building within the community. 

Additionally, \change{the cameras on the AR HMD may pose privacy issues, especially in paramedic tasks}. P1 explained his concern with wearing AR glasses when interacting with patients: \textit{``If you think about the medical situations that we go to, it's their worst moment in their life. So if they're looking at you and going like: `What the heck are they doing here? I can't trust them.' This defeats the purpose [of first responders].''} 
As a result, P3 proposed integrating a small heads-up display to FR's face shield to support navigation in ambulances driving and firefighting, while \change{paramedic P6} suggested integrating only an intelligent voice agent in the face shield for necessary support \change{for privacy purpose}. 


\change{\subsection{AR Maps for Spatial Awareness}\label{AR_map}} 


\change{Maps that provide structural representations of a geographical area serve as critical tools in developing and sustaining spatial awareness during in-the-field tasks \cite{nunes2017augmented, campos2019mobile, grandi_2021_Design}. We compare the 2D and 3D maps on the AR HMD with FRs.

\subsubsection{2D vs. 3D Maps.} Participants found both 2D and 3D maps useful for maintaining spatial awareness.  
However, the detail level and complexity of the different map design affected FR's preferences and led to different usage scenarios:

\textbf{\textit{2D Maps for Urgent and Hazardous Scenarios.}}
Compared to 2D maps that simply show the top view of the space, the 3D map presents more complex details and requires more interactions (e.g., rotate the map to see different perspectives), which can be less practical in urgent scenarios (P4, P9).
Sixteen participants preferred 2D maps over 3D, appreciating the simple yet sufficient information provided under emergency, such as search and rescue (P4, P5, P7, P10). 
As P5 explained: \textit{``We prioritize searching bedrooms and livable spaces where people are commonly found. [With the 2D map], we quickly look at it and see where the bedrooms are at, versus having to go through the entire house room by room.''} Two participants (P1, P19) 
emphasized the needs for labeling their current position on the 2D map for self-orientation: \textit{``In disaster scenarios, if I can have a layout of a building that's been bombed to it, and I can see where I'm at in accordance with the layout, that will be huge and helpful''} (P1). 

\textbf{\textit{3D Maps for Training and Pre-planning.}}
In contrast, the 3D map was considered as more suitable for training (P3, P4, P8, P10), particularly for demonstrating the flammable objects (P8) and fire development (P10), since it allowed FRs to see the space from multiple angles. P3 and P7 also noted the potential of 3D maps for pre-planning due to its rich details and adjustable viewing angles: \textit{``We'll go into a house or a building that might have an interesting or strange layout or something worth considering. And if you can pull up [the 3D map] and say, `This is why it's weird.' And you could turn it around and say, `Oh the couches are over here.' I think the 3D map would be awesome''} (P3).
Moreover, P4 noted that 3D maps would benefit incident commanders more than crew members to thoroughly monitor all activities in the field: \textit{``From a chief's perspective though, that 3D map would probably help them a ton...especially if they could have a head overlay for them [showing] where each individuals were at in the structure.''}}





\subsubsection{\change{``Multi-Layered Map'' to Balance Simplicity and Details}}
\change{While all participants acknowledged the critical role of maps in emergency response, the specific information needed on the map varied across FR roles. Therefore, the map design should meet the needs of different FRs while avoiding information overloading in emergency (P4, P9).} To achieve this goal, participants suggested a \textbf{multi-layer map design} that \change{presents commonly-needed \textbf{base information} with \textbf{switchable layers for on-demand information} required by different FR roles}. As P10 envisioned: \change{\textit{``If there would be an ability to put layers on the map, so if there were known hazards materials, it would be nice to be able to say: `Okay, I'm just going to pull up the hazardous materials layer.' If there's nothing there, then I'm just going to wipe that out.''}}


\textbf{\textit{\change{Streamlined Base Map with} Structural Elements.}} Two participants (P6, P10) suggested that an AR map should have a base layer that consists only the most fundamental, static, and structural elements to ensure information accuracy \change{and consistency for all FRs}, mitigating confusions caused by dynamic changes.  For example, movable furniture should not be involved in the base layer. As P6 explained, \textit{``I think having furniture and stuff on the map could be a hindrance to a degree, because that's something changing from tenant to tenant or day to day. So it would almost be one of those things where you expect it to be somewhere, and then it isn't where it was on the map...So I think a structural unchanging layout would be more valuable. You would make the map assembly once and that would be it.''} Examples of structural elements include walls and any openings in a space, such as windows and exits, as suggested by P6.

\textbf{\textit{Switchable Layers for On-demand Information.}} Beyond the base layer, different FR roles desired to track different dynamic information on the map. \change{For example, participants suggested to label all egress points (P1, P6, P14, P17, P19), rooms that had been searched, and areas with toxic gas or extreme high temperate (P3, P8, P9, P10) for firefighters}; hidden hazards and teammate locations for firefighters and LE (P6, P21, P24); \change{suspect locations for the LE (P15)}; and victims or \textit{``warm bodies''} for paramedics (P15). 
\change{Due to the role-specific needs}, participants wanted such context-dependent information to be presented on separated layers, which can be flexibly toggled on and off for different FRs and tasks.
As P6 mentioned: \textit{``It's possible to have different modes that [contain] certain features [to be] turned on and off. So let's say you have a search mode, it turns on an array of options, [where] you have door and window detection, you've got organic bodies protection like humans or pets. So those things were all turned on. 
And all [different information] could probably be situated within a mode framework.''} 

\subsection{Directional Cues for Different Navigation Contexts} \label{navigation} 
\change{Unlike maps that provide a high-level overview of the space, directional cues (e.g., arrows, path) offer precise, world-anchored guidance \cite{nunes2017augmented, Nunes_2018_ARapp}. Two participants (P10, P15) found that the world-anchored directional cues provided first-person view immersed in the physical environment and enabled them to quickly and accurately gauge distances while moving. 
P15 used driving ambulance as an example to compare map and the world-anchored directional cues: \textit{``On a map, yes, you can see how far away your turn is. But how do I know how far [it] is in reality? Because when you zoom in and out, it's different distance. But if you're actually looking straight ahead, and you see it's a long path, you're like, `Okay, I got a ways to go.' The path gets shorter and shorter, and you know where you are supposed to turn. You also don't have to keep looking at the computer and back to the road.''}

While all participants considered the directional cues highly beneficial, their needs and design preferences differed according to their unique working scenarios and roles: }


\textbf{\textit{\change{Back-tracing in Visually Challenging Environments for Firefighters.}}}
\change{Firefighters often operate in more visually challenging environments with dense smoke than EMS and LE \cite{grandi_2021_Design}, thus being easier to get lost in the fire (P3, P11, P15). Firefighter P11 explained their unique navigation challenge: \textit{``The majority of firefighter fatalities are incapacitation in a smoke filled environment. They can't find their way out, and they run out of air. In a panic, [they] take the face piece off, and typically there is smoke inhalation.''} Although they used traditional methods, such as rope, hose lines, or counting walls to track the environment (P7, P11, P13, P21), 
these approaches were not consistently reliable (P7, P11), 
as P7 pointed out: \textit{``Rope can be cut, rope can be lost, and it can be tangled.''}}

To mitigate such risks, three participants (P3, P6, P7) desired to mark their path once entered the incident scene for back-tracing. For example, P3 proposed a digital breadcrumbs design: \textit{``In a firefighting scenario, if AR cue was able to track you via a digital breadcrumb, and if you get turned around, you can look up, and it would show you the way out, say like ‘Hey, this is the way you came’, that would feel pretty good.''} \change{Another benefit of the directional cues head-mounted AR was to free the firefighters' hand from holding the rope: \textit{``We have to have a hand on the rope. If this is showing us where our exit is at all times, now we have two hands available to do our search. This would be much more efficient'' (P11).}}




\textbf{\textit{Subtle Cues for \change{Emergency Vehicle Driving.}}}
\change{Different from other FR roles, paramedics' (P7, P17) main responsibilities involve driving ambulances and transporting patients efficiently between incident scenes and emergency rooms. As opposed to navigating in a dark, smokey environment as a firefighter, driving emergency vehicles is a less physically intense but equally demanding challenge, requiring FRs to acutely focus and adapt to the rapidly changing surroundings (P5, P7, P13, P16, P15). As such, instead of obtrusive cues to attract users' attention, the directional cues for emergency vehicle driving needs to be subtle, not distracting the drivers from focusing on the road. 
As P5 explained the difference: \textit{``If I need these arrows [during firefighting], that would be pretty serious situation. I'm probably gonna be scrambling and I want the arrows to be very obvious. [However,] for driving situation, you do want it [to be] pretty low-key ... something that we can still pay attention to the traffic, or something pretty benign and off to the side, because you should know where you're going.''}}



\change{\textbf{\textit{Real-time Marking for Collaboration.}}
FRs often navigate unfamiliar structures that they have never been to and the navigation can become more difficult due to outdated buildings blueprints (P10). In such scenarios, leveraging the prior knowledge from teammates becomes essential. As P13 explained: \textit{``You're relying on the knowledge of the people that have been at that building prior or over time. So [instead of] going through all the building every time, it would be better to have historic knowledge that everyone can have access to.''} As such, four participants (P1, P9, P12, P13) proposed the idea of real-time arrow placement in the physical environment, enabling other team members to follow the route that was already explored. For instance, P1 envisioned how the real-time arrow placement could facilitate efficient collaboration than verbally describing the scene to teammates: \textit{`` So let's say that you're going into search somewhere, obviously, there's a disaster. [With] the arrow placement, I don't need to explain to the next search crew how to get there. And it would be like, 'just follow the arrows that I placed.' It's a lot easier than trying to describe [a scene], such as trying left at the red bar, go next to the fire, and then go to the left with a broken car.''}}

\textbf{\textit{AR Compass for Outdoor Navigation.}} Besides indoor navigation, P6
brought up the need for real-time self-orientation in outdoor navigation when operating outside of the incident building, such as locating Fire Department Connection. He proposed a virtual compass in AR: \textit{``I can envision the use of compass. So from my perspective in the HoloLens 2, I can see it somewhat on the floor, so then north is stays here. And then the compass kind of changes as you move.''} \change{This echoed a design in prior work \cite{Nunes_2018_ARapp} that a compass was integrated into a mobile AR interface. However, our participant preferred a more immersive design on AR HMD, where the AR compass was placed on the ground and centered at the user's position for a first-person perspective.}

\subsection{Highlighting Cues for Objects of Interests} \label{highlighting}
\change{Highlighting cues complement maps and directional cues by not only enhancing objects of interests but also providing information on how to approach them effectively (P5, P10). 
Multiple participants (e.g., P3, P7, P15) emphasized that the highlighting cues enabled them to understand the size and shape of important objects and better determine subsequent actions. We report how FRs preferred different highlighting cues and what types of elements they wanted to highlight.

\subsubsection{Outlines vs. Shaded Overlay.}
By comparing the outline and shaded overlay (Figure \ref{fig:probes}C-D), we found that FRs had different preferences between the two cues. Specifically, twelve firefighters predominantly preferred highlighting egress points using outlines, such as doors (P1, P2, P5, P10), windows (P1, P3, P23), and \textit{``any kinds of holes''} (P5). They found the solid lines and high-contrast color of outlines allowed them to accurately locate exits/entrances and their exact boundaries in low visibility environment, which made it \textit{``unmissable and unmistakable''} (P3). Conversely, the shade overlay was less favored as it obscured details, hindering their common tasks like forcible door entry. As P3 noted: \textit{``My initial reaction to the shade is that it obstructs the detail. I might want to see what type of door is that? What handle? What locking mechanism? Is it a metal door or a wood door? With this outline, I can see those very clearly. It's showing me that [the object is] there, but it's not getting in the way of gathering [real-world] information.''} 

Unlike firefighters who preferred highlighting entrance and exits, EMS and LE desired to highlight smaller and less visible or detectable objects. For example, paramedics P13 wished to identify the implantable medical devices (e.g., pacemaker) for more efficient patient assessment, and LE officer P18 wanted to identify any weapons carried by suspects. For those small and hidden objects, participants (P7, P18) found shaded overlay that \textit{``lit up''} the entire objects were more eye-catching.}


\subsubsection{Highlighting beyond the Whole Objects.} Besides augmenting an entire object with the highlighting cues, the FRs also indicated the specific elements on an object or properties of an object to highlight. 

\textit{\textbf{\change{Outlines Informing Object Affordance.}
}} Highlighting object affordance---the interactive components on an object and what interaction is supported by the object---was shown to be helpful for most participants across different FR roles. Take the door as an example, participants wanted \textbf{AR cues to highlight the door hinge, knob, and swinging direction to quickly determine the door forcing strategy} (P4, P8, P10, P11, P12).  
As firefighter P10 explained: \textit{``Depending on how the door swings, whether it swings on the left side or the right side, if it swings in or it swings out from a couple of different standpoints, [we decide how to] force the door open.''} Knowing the door affordance can also help FRs select a safe standing position by the door. 
P10 explained the selection of standing position from the paramedic perspective: \textit{``So if I'm going to do an EMS call, where there's maybe somebody was involved in a fight, I always approach the door like there's potential something harmful on the other side. So when you stand at a door, you want to stand on the side with the doorknob. Because if someone intends you harm from the backside and I'm standing on the hinge side, they open that up and have a direct line of sight to you. [But] when I'm standing on the doorknob side, they have to open the door up in order to get around.''} 

\textbf{\textit{\change{Patterned Shades Indicating Object Properties.}
}} \change{Both FF and LE offices expressed the needs} of knowing key properties of an object, such as the material (P4, P6) and the temperature (P6, P14). As P6 elaborated: \textit{``Material matters at how we would force a door open. If you are unable to see the door visually, but knowing that it's a metal door, you would definitely approach that in a different way [than not knowing it beforehand]... you will use different tools and different ways to approach the door.''} Instead of a plain shade, P6 suggested using overlay with patterns, such as a \textit{``honeycomb or cube''} pattern on the door, to visualize the materials.

Object temperature was especially vital for firefighters (P9). P6 and P14 proposed to \textbf{\change{combine the imaging from a thermal camera with the highlighting cue, thus rending the thermal gradient as an AR overlay}} to highlight the heat distribution on physical surfaces. As P14 indicated: \textit{``[A way] of incorporating our Thermal Imaging Camera into [the AR glasses] is using a gradation pattern, showing us the difference between hot and cold and the intensity of it.''} Such integration would be intuitive to firefighters since the thermal camera was the only technology they used in the field. 




\subsection{Labeling Cues for Hidden or Detailed Information} \label{labeling}
\change{As opposed to highlighting cues that depict the shape of the objects, labeling cues (e.g., icons, text) are used to label the position of objects and provide more detailed information \cite{Nunes_2018_ARapp, siu2013sidebars, Weichelt_2018_Augmented}. 
}
All participants wanted to \textbf{use the labeling cues to label hidden or small hazards}. Firefighters, in particular, highlighted the risk associated with concealed fire sources, such as floor holes and void spaces: \textit{``[Those] older buildings got remodeled and have tons of void spaces, and that is both a hazard and a huge spot for fires to grow and remain undetected''} (P3). \change{Moreover, firefighters often needed to mark checked rooms during search and rescue tasks (P4, P10), and they found labeling cues in AR could avoid destroying properties due to physically marking and altering the environment: \textit{``The fear [of marking checked rooms] right now is that if we market it with spray paint or something, we're ruining that door. So with an AR, we'd be like, label the room that has been searched. We're not causing any damage to that door, and [teammates will] know that they don't need to search that room.'' (P16).}}
For EMS and LE, smaller hazards, such as used needles, shattered glasses (P7), weapons, and open containers for drunk drivers identification (P18), were less detectable and thus needed to be labeled. To label hazards properly, P17 suggested that \textbf{the AR icons should follow commonly agreed standards}, such as \textit{DOT CHART 17}\footnote{DOT CHART 17. \url{https://www.phmsa.dot.gov/sites/phmsa.dot.gov/files/2022-10/Chart17-10-06-2022-508-REM.pdf}}, a guide for Hazardous Materials Markings, Labeling, and Placarding. 



Two participants (P4, P6) believed that the icons can be used to convey the attributes of objects, human, and FR equipment. For example, P4 suggesting using the size of an icon to indicate a victim's weight to determine the needs for rescuing resources: \textit{``A component that could really be helpful with the icon would be if it could distinguish a size. Because even a very small person can take a lot of resources to get out. And if it's not a small person, then it can give the preemptive call out that we would need more help.''} As opposed to world-anchored icons that label physical environment, some participants (P3, P6, P20) desired a head-attached gauge icon to indicate their equipment information, such as the remaining air in the firefighter SCBA mask, informing the distance they could travel before having to return. 

\change{Compared to icons, participants (P1, P15, P18) suggested using text to label more detailed information, such as the identity of all on-site human subjects. Since tasks in the field usually involve multiple parties (e.g., victims, victim's family members, potential suspects), labeling their detailed identity would facilitate situation awareness and ensure FRs' safety (P1, P15, P18). As P15 explained: \textit{``For example, there could be a kid over there, a neighbor there, and this is mom and that is dad, who is a possible abuser...I don't know all that information. If they could be relayed without talking or interpreting it [and] everyone has the same information, I could see it being really helpful [for] a big safety issue.''}}

\subsection{Preferences on Properties of AR Cues} \label{properties}
Participants discussed on four properties of the AR cues to detail their preferred design forms, including colors, thickness, opacity, and placement.

\textbf{\textit{Colors.}}
\change{In mass casualty incidents that involve multiple tasks or situations in different urgency levels, FRs must quickly assess the scene and prioritize the most critical situations \cite{Nelson_2022_explore, killeen2006}. Six participants (P1, P2, P15, P20, P22, P24) 
found color of AR cues was helpful to indicate the severity level of an incident, and paramedics also mentioned that colors were commonly used for triage (P7, P15). 
In terms of the selection of colors, some participants suggested} \textit{``following standard traffic lights''} to indicate the hazard status (P4, P15, P24), where red representing the most dangerous condition that requires FRs' immediate attention. On the contrary, three participants (P7, P9, P21) preferred not using colors commonly seen in emergency response, such as red and orange, to prevent confusion because \textit{``a lot of our stuff is already in red''} (firefighter P7). 
As a result, P7 and P9 suggested to adopt \textit{``unnatural but bright colors,''} such as cyan or magenta, to make the AR cues stand out from the background. 

\change{Importantly, as opposed to cues on video see-through devices (e.g., mobile AR), three participants (P1, P3, P7) 
found \textbf{the perceived color of visual cues on the optical see-through AR HMD could dynamically change with the physical environment} (e.g., different lighting condition).} 
Nine participant emphasized the importance of color visibility against the real-world background, considering the diverse environments where FRs operated. They suggested that the AR cues should adapt to the background colors to ensure high contrast. As P7 illustrated: \textit{``[The color] will depend on the space. [In the case of] water rescue and hypothermia [...or] if we're going to rescue someone who had fell in the lake, and we're on acres of snowy land, [these green cues] could be hard to distinguish.''} 
P3 further suggested avoiding nuances in colors since they were barely detectable in low visibility environments \change{that firefighters often involved in.} 


\textbf{\textit{Thickness and Opacity.}}
\change{Similarly,} thickness and opacity of the AR cues were also suggested to indicate emergency levels. For example, P15 preferred to label patients with more urgent demands using thicker contours or opaque labels, so that the more urgent patient would attract more attention from the paramedics and the less urgent points of interest should \textit{``fade into the background.''} 
\change{However, unlike color selection where visibility mattered most, \textbf{the usage of thickness and opacity must balance visibility and intrusiveness}, providing clearly visible guidance while not blocking the physical surroundings and interfering with FRs' perception (P14, P15). This was particularly important for emergency vehicles driving tasks with dynamically changing environments. As P15 illustrated: \textit{``I'm looking at the arrow, but I'm also noticing that there is a car coming at me behind the arrow. It could distract me like `Is that car actually coming at me?'''}
}

\textbf{\textit{Placement.}}
\change{Participants expressed different placement preferences for different types of AR cues. For labeling cues, multiple participants (e.g., P4, P11, P16) preferred to place the icons \textit{``right on the particular object itself''} (P16). Because they found the cue placement was an important indicator of an object's exact position, guiding them to locate that object accurately. On the contrary,} six participants preferred to attach labeling cues to the side of an object of interests but not covering it, allowing FRs to see the original object easily. 

\change{For directional cues, however, firefighters (P9, P10, P11, P13) strongly preferred attaching them to the lower ground for reachability. As P9 elaborated on their practices in smoke-filled environments: \textit{``When we have no sense of direction in the dark and [the environment] are super hot, we try to stay as low as possible, and looking around would not be an option.''}. As such, firefighters suggested that the placement of directional cues should follow their navigation practices: \textit{``People are on the floor, that's what we are trained for ... so I would bring those arrows down below. [When we are on the floor,] 
we're either [navigating] on our bellies, possibly as high as on our knees, but more so on our bellies'' (P11).}} 


\subsection{Multimodal Interactions in the Field} \label{interactions}

While HoloLens 2 supported multi-modal interactions, speech command was considered the most promising interaction technique by eight participants. As FRs’ hands were usually occupied by heavy gloves, tools, and environment exploration, eight participants worried that the hand recognition accuracy would be reduced \change{for in-the-air gestures}. Three (P3, P6, P11) participants found the eye tracking distracting, forcing their gaze (or attention) away from the primary task. We report participants' needs for different interaction techniques in the field. 




\textbf{\textit{Short, Distinctive Speech Commands.}}
\change{Speech command was considered as the most promising interaction technique by six participants. All types of FRs valued its hands-free nature (e.g., P10, P15, P18), as they often carried multiple equipment and had both hands occupied. For firefighters worked in extreme conditions, hands-free interactions even became a must-have, as P1 indicated: \textit{``I've been in tunnels where I can't really move my hands, they're out in front or behind.''}} However, since audio was the main communication channel for FRs in the field, four participants (P3, P5, P19, P21) emphasized that the speech commands had to be \textit{``a pretty specific phrasing to bring [a function] up''} (P3), distinguishing from team communication. More importantly, three participants (P1, P4, P6) highlighted that the speech commands must be short for memorability and safety because firefighters had limited air with SCBA masks. 

\textbf{\textit{\change{Seamless Gaze Interaction.}}} 
\change{Gaze input was another hands-free interaction uniquely afforded by head-mounted AR and favored by eight participants. Unlike voice interaction, which may not be recognizable in noisy environments (P5, P6, P9, P10), gaze required minimal effort to interact with cues. As P15 described: \textit{``You just looking at it. You don't have to do anything.''} P12 also found that \textit{``it would be much faster to gaze than touch during during an emergency situation.''} However, three participants (P3, P6, P11) were concerned that moving their gaze away from the task they were focusing on for interaction would be a distraction. \textbf{Firefighter P3 questioned the usability of gaze input in low visibility environments}: \textit{``I feel like your eyes just kind of wandering around when you don't have something to fixate on. In my experience, I'm always looking around even if I can't see anything. When you're in a fire, you're constantly scanning for information, especially the less information you have available to you in front of you, the more you're trying to gather.''}}

\textbf{\textit{Tangible Control.}}
P6 suggested adding such physical buttons to provide tangible assurance in complex situations with various rapidly changing and distracting elements: \textit{``some sort of mechanism where you could feel that you're doing it [were] needed...so a button on the side [that] is very big, [so that] gloved hand can grip the button, and there's a click, a feedback that we can feel.''} \change{While HoloLens 2' typical interactions do not involve physical buttons, it provides several buttons to turn on and off the device and adjust audio volume. These buttons could be expanded as input methods for FRs to provide tangible feedback if necessary.}  


\textbf{\textit{Intense Head Vibration for Extreme Dangers.}}
Three participants (P6, P7, P17) desired haptic alert from the AR HMD. Such alert should be seldom-triggered but highly powerful to attract FRs' immediate attention to extreme conditions. For example, the firefighters (P6, P17) suggested using a strong vibration from the HMD to indicate the Mayday situations, where the life of a firefighter or somebody they are with is threatened. As detailed by P6: \textit{``I don't think I want my head rattled. But under an extreme condition, if there was a mayday, it makes sure everybody was listening.''} For LE and EMS, the extreme conditions referred to suspicious person carrying life threatening weapons (P7). 

\section{Discussion}
\change{
We contributed an in-depth exploration into the design space of optical see-through AR HMDs to support FRs in the field. Through a user-centered approach, we provided FRs first-hand experiences with the most state-of-the-art AR HMD---HoloLens 2---to understand their needs and concerns. We also involved diverse FR roles (FF, LE, EMS) who often collaborate in the field to understand their common or unique needs, thus inspiring AR HMD design that can be connected and shared by different FR members. 

Our findings answered three research questions (in Introduction) from the perspectives of hardware, interactions, and AR feedback. While enabling comfortable wearing for most FRs (except for one female FR), AR HMDs for FRs should be wraparound instead of semi-enclosed to provide protection and wide field-of-view (RQ1). 
Among the various interactions supported by HoloLens 2, we revealed that speech commands were favored by all FRs, enabling hands-free operation with minimal distraction, while other modalities could be used to compensate for certain scenarios, such as gaze input in noisy environment or vibration alert on the HMD in extremely dangerous situation (RQ2). 
In terms of AR feedback, we solicited role-specific and scenario-specific needs among different FRs, identifying suitable use cases and design forms for different types of AR cues (RQ3). For example, to support consistent space awareness, a multi-layer map design is preferred that has the structural layout as the base layer shared by all FRs as well as switchable layers for on-demand, role-specific information controlled by different FRs (Section \ref{AR_map}). Obvious directional cues are needed for back-tracing for firefighters in visually challenging environments, while more subtle direction cues should be used in emergency vehicle driving to avoid visual blocking (Section\ref{navigation}). For highlighting and labeling cues, we found that outline highlighting were preferred by FF and LE officers to enhance properties and affordance of essential on-site objects, while shaded overlays, icons, and text were preferred to  label hidden or small hazards with more detailed information (Section \ref{highlighting}-\ref{labeling}). 


In this section, we derive design guidelines, including both generic guidelines and role/scenario-specific guidelines, to inform practical and safe use of AR HMDs in the field.}

\subsection{Design Guidelines for Head-mounted AR Systems in Emergency Response}
Building upon our findings and prior literature \cite{hand_recog, kim2019foveated, voice_recog, ParkKim2020NoiseCancellation, jiang2004siren, Nunes_2018_ARapp}, we derive AR design guidelines from the aspects of hardware, interaction techniques, and AR feedback, \change{that consider different FR roles and in-field tasks}: 

\change{
\subsubsection{Hardware Design (G1-G3)} \hfill\\ 

\textbf{G1. Optimize Comfort and Safety in Ergonomics Design}
        \begin{itemize}
            \item \textbf{\textit{G1.1. The form factor of AR HMDs should consider gender-specific needs}} (P15). The weight should be suitable for female FRs who may have lower weight tolerance than their male counterparts. Women also generally have smaller head sizes and different facial structures (e.g., lower nose bridge, fuller cheekbones) compared to men, requiring adjustments to ensure a comfortable fit and effective sealing. 
            \item \textbf{\textit{G1.2. AR HMDs need to enable adjustable tightness across different in-field scenarios}} (P1). In-field tasks involve different levels of movements, thus requiring different HMD wearing stability. Too tight attachment to head could reduce comfort, however, certain tasks (e.g., tunnel rescues by firefighters) require large body movements so that the HMDs have to be additionally tight and stable. As such, flexible tightness adjustments, ranging from stable wear in routine tasks to extreme tightness for long-time wearing in physically heavy tasks, need to be supported. 
            \item \textbf{\textit{G1.3. The hardware design must follow established safety standards in emergency response}} (P1, P3, P16). Given the high-risk nature of in-field tasks, AR HMDs should fully conceal FRs' eye area for multiple hazards prevention (e.g., smoke, blood splatter), following PPE standards in the field. 
            \item \textbf{\textit{G1.4. Head-mounted AR should provide wide field of view for full situation awareness}} (P1, P10, P18). Current OST HMDs have small field of view (e.g., HoloLens 2 has a 50\textdegree ~horizontal field of view \cite{narrow_degree}), largely limiting the notability of world-anchored virtual augmentations and the capability to enhance FRs' situational awareness. While extending displays to cover peripheral vision would be optimal, virtual indicators should also be considered to present off-screen information \cite{kim2019foveated}. 
        \end{itemize}
       
    \textbf{G2. Employ Socially Acceptable Form Factor}
        \begin{itemize}
            \item \textbf{\textit{G2.1. The HMD design should follow common social etiquette by showing FRs' face or eyes to mitigate victims' or patients' confusion and fear}} (P8). This is specifically important for EMS who need to build trust with patients and victims. For example, increasing the transparency of visors in AR HMDs to facilitate eye contact, and reducing the size of HMDs' front enclosure to reveal more facial expressions. However, this guideline should be considered together with the safety standards (G1.3). As such, a fully concealed HMD with transparent display or outward display (e.g., Apple Vision Pro \cite{apple_vision_pro}) showing the wearers' face should be considered. 
            \item \textbf{\textit{G2.2. Camera integration should consider privacy concerns}} (P1, P6). Given the long-standing use and public acceptance of body-worn cameras in LE \cite{McCamman2017PoliceBodyCameras}, AR HMDs could adopt similar standards for their cameras integration. This includes minimizing the camera's physical presence while ensuring that individuals are informed when the camera is actively recording. For example, one typical method is using an LED light to indicate the camera usage \cite{Portnoff_2015_watching}, and the HoloLens 2 has incorporated such a light indicator on its visor for privacy protection. However, this feature could potentially distract FRs from their tasks and interfere with their vision, especially in low-visibility environments (P3). To address these, the privacy notification design on AR HMDs must consider the needs of both the public and FRs, informing camera usage without interrupting FRs' work. 
        \end{itemize}
        
    \textbf{G3. Ensure Durability in Extreme Conditions}
        \begin{itemize}
            \item \textbf{\textit{G3.1. AR HMDs need to be robust in extreme conditions, particularly for FFs who work in extreme heat}} (P1, P4). While the Hololens 2 is not well suited for high-temperature firefighting tasks, researchers and practitioners have started developing heat-tolerant AR devices (e.g., the Fusion Vision System, a helmet mounted AR device that can be attached to FF's helmet \cite{LonganVision2023}) for such extreme scenarios. Future AR HMDs should incorporate comparable heat-resistant technology for robust use in the field. 
            \item \textbf{\textit{G3.2. AR HMDs must fail gracefully in non-operational situations}} (P9, P10). While all technology has reliability issues, FR-centered technology must ensure that technology failure returns users to a not-worse-than-baseline status, such as when batteries die. As such, OST AR with transparent displays would be recommended. 
        \end{itemize}

\subsubsection{Interaction Techniques (G4-G5)}\hfill\\

\textbf{G4. Support Accurate, Robust Input in Various In-Field Environments}
        \begin{itemize}
            \item \textbf{\textit{G4.1. In-the-air gesture recognition should consider in-field needs}} (P17). In-the-air gestures is a major interaction for HoloLens 2. However, in-field tasks can involve low-visibility environments, hands with thick gloves, and occupied hands by tools, which can potentially affect in-the-air gesture recognition. From the hardware aspect, incorporating more sensing capability (e.g., thermal camera) can potentially enhance the recognition accuracy \cite{hand_recog}. 
            From the gesture design aspects, fine-grained gestures involving fingers or nuanced hand movements should be avoided, as FFs often wear bulky gloves in the field.
            \item \textbf{\textit{G4.2. The speech commands must consider noisy environments}} (P5, P6, P9, P10). Many in-field tasks, especially in mass-casualty incidents, involve noisy environment that may reduce speech recognition capability. An improvement consider noise-cancellation technology with voice identification algorithms to distinguish the FRs' voice from background noise \cite{voice_recog, ParkKim2020NoiseCancellation}. The speech commands should also be carefully designed to avoid conflicts with FRs' verbal communication with teammates. 
            \item \textbf{\textit{G4.3. The gaze interaction should consider low-visibility conditions}} (P3). Low visibility environments where FFs work in could be challenging for gaze interactions due to both recognition issues and fixation issues. For recognition issues, gaze estimation algorithms in low-light conditions \cite{kim2020gaze} should be integrated to AR HMDs. For fixation issues, since FFs are concerned about eye wandering in dark environments without targets to focus on, suitable object augmentations should be generated  based on recognition on depth or thermal images to provide fixation targets for the FFs for further gaze interaction. 
        \end{itemize}
        
    \textbf{G5. Avoid Attention-demanding Interactions, Except for Extreme Situations}
        \begin{itemize}
            \item \textbf{\textit{G5.1. The interactions on AR HMDs should provide minimal distractions to allow FRs to focus on their primary tasks}} (P3, P6, P11). Speech commands should be intuitive and straightforward, causing little to no cognitive loads. Hand gestures should not interfere with primary actions, such as holding the rope by FFs. Gaze interaction should not deviate FRs' attention from their primary tasks. 
            \item \textbf{\textit{G5.2. Haptic feedback on HMDs could be considered as an invasive, attention-catching alert for extreme situations}} (P6, P7, P17). With the integration of haptic components (e.g., vibration) attached to the HMD, haptic feedback on head can be used to attract FRs' immediate attention in extreme conditions. For example, the strength and frequency of vibrations can be used to represent various urgency levels of incidents. Prior research has explored the haptic sensitivity on users' head area \cite{thermor_vr}, which could inspire the haptic components integration. 
        \end{itemize}

\subsubsection{AR Feedback Design (G6-G8)} \hfill\\

    \textbf{G6. Meet Universal Preferences}
        \begin{itemize}
            \item \textbf{\textit{G6.1. AR cues should avoid conflict or confusion with real world information}} (P7). To avoid confusion between AR cues with physical-world objects, the realistic-looking cues or properties with certain meanings should only be used very intentionally. For example, avoid using red-color cues in firefighting scenes since it is a symolic color that are frequently used to label FF equipment.
            \item \textbf{\textit{G6.2. AR cues should not obscure large sections of the physical world}} 
            (P5, P7, P15).  FRs need to prioritize real-world environment for situational awareness, with the AR HMDs and other technologies as supplementary tools instead of relying on them. For example, leveraging AR HMDs to provide subtle directional cue anchored to the floor rather than floating in the air or attaching to the wall. 
            \item \textbf{\textit{G6.3 Augmenting affordance rather than the entire object}} (P4, P8, P10, P11, P12). As opposed to the whole object of interest, the interactable component on the object and how to interact with it (i.e., affordance) would be more important and need immediate attention, especially in emergency response. For example, instead of highlighting the entire door, augmenting the door knob and the opening direction of the door could be more important. State-of-the-art affordance detection models \cite{do2018affordancenet} could be integrated to support this need. 
        \end{itemize}
        
    \textbf{G7. Situate Design for Scenario-specific Needs}
        \begin{itemize}
            \item \textbf{\textit{G7.1. AR cues should be adaptive or adaptable to the emergency level of an incident}} (P4, P9, P10). For instance, overview cues (e.g., maps) should provide FRs the most essential information from a glance in emergency, while it could contain more details and allow more interactions in routine tasks or training.
            \item \textbf{\textit{G7.2. AR cues must dynamically adapt to the physical environments to ensure visibility}} (P1, P3, P7). On OST HMDs, the visibility of AR cues can be significantly affected by the real-world background, such as the background color, texture, and lighting condition. Since FRs' real-world background could constantly change in emergency, the appearance of AR cues (e.g., color, brightness, transparency) should adapt to the background in real time to both ensure visibility and avoid visual overwhelmedness.  
            \item \textbf{\textit{G7.3. The properties of AR cues should consider FRs' pace of movement}} (P5, P7, P13, P16, P15). In fast moving scenarios such as driving an ambulance, the properties (e.g., size, placement) of AR cues should be designed differently from slow moving tasks such as foot patrol.        
            For example, when driving an ambulance, subtle cues should be considered to avoid blocking any real world information. 
        \end{itemize} 

    \textbf{G8. Tailor AR Cues to Role-specific Needs}
        \begin{itemize}
            \item \textbf{\textit{G8.1. Prioritizing self-tracking to support FF navigation}} (P3, P11, P15). With the unique challenge of navigating in visually challenging environment, both overview and directional cues should be leveraged to support navigation from different perspectives. Specifically, overview cues should pinpoint FF's real-time position in the map for universal awareness, and directional cues should back-trace FF's path for situational awareness. 
            \item \textbf{\textit{G8.2. Attaching AR cues to the ground for FF's reachability}} (P9, P10, P11, P13). Since FFs are trained to navigate and work on lower ground in smoke-filled environment, AR cues should be positioned on the floor level so that they do not need to spend extra efforts to search or locate cues in the environment. 
            \item \textbf{\textit{G8.3. Highlighting the shape and properties of hazards for LE and EMS}} (P13, P18). LE and EMS usually need to pay attention to small or even hidden objects (e.g., weapon, used needles). For these small objects, labeling cues (e.g., icons, text) and shaded overlay on the objects to depict its appearance and detailed information are preferred to enable LE and EMS to collect more information and plan actions quickly. 
        \end{itemize}
        

}

\subsection{Limitations \& Future Work}
Our research has limitations. While presenting FRs first-hand AR experiences, they explored the technology in a relatively ideal lab environment, \change{with eight watching videos of the AR cues}. Their preferences may differ if using the AR glasses in the field in the real world. The challenging environment in the field may further hinder the capability of the AR HMDs, such as reducing or disabling recognition capability. In the future, more empirical studies should be conducted to collect in-situ evidence and validate the guidelines derived from current qualitative study, such as quantitatively measuring FRs' performance using AR HMDs in different in-field tasks. Moreover, our participant distribution has a strong gender bias towards male FRs, which align with the current FR gender distribution \cite{gender_1, gender_2}. 
However, female users may have unique technology preferences, such as the weight and form factor of the hardware. \change{For example, one female FR P15 found the HoloLens 2 heavy and not suitable for long-term use.} It is equally important to design suitable technologies for female FRs. As such, our future research will expand the participant diversity by involving FRs with different identities and background, such as female, FRs who are color blind, to ensure a comprehensive coverage. \change{Finally, our participant distribution skews towards firefighters, leading to more firefighter-specific findings. Future research should also considering expanding the recruitment for other disciplines}.


\section{Conclusion}

In this paper, we thoroughly explored the design space of optical-see-through AR to support in-field tasks for FRs by providing them first-hand experiences with a state-of-the-art AR HMD and four representative AR cues. The study covered FRs in different roles (i.e., FF, LE, EMS) and investigated their unique needs in different emergency response scenarios. Our findings revealed both generic and role-specific preferences, needs, and concerns of FRs from three aspects: hardware form factors, interaction techniques, and AR feedback. 
We finally derived actionable guidelines to inspire the design and development of AR HMD technology that supports in-field emergency response and collaboration among various FRs. 

\begin{acks}
We thank all anonymous participants for their efforts and valuable feedback. This work was partially supported by an award from the U.S. Department of Commerce \#70NANB21H043.
\end{acks}

\bibliographystyle{ACM-Reference-Format}
\bibliography{sources.bib}

\appendix
\section{Appendix}

\subsection{Pre-screening Questionnaire}
Participants were eligible to participate in our study if they were 18 years old or above, available for an in-person study at our lab, and had experiences working as a first responder. 
\begin{enumerate}
    \item Are you 18+ years old?
    \item Do you live in XX, XX? (If yes write in yes, if no write the city you live in.)
    \item What is your current occupation?
    \item Are you currently a first responder or have been one in the past?
    \item Please specify your years of experiences as a first responder with start and end years (e.g., 2010 - 2015; 2020 - present; 3 months). 
    \item Do you have any prior experiences with AR glasses?
    \item Please describe your experiences with AR glasses if you have any (e.g., type of AR glasses, AR application you used before, what do you use it for); enter N/A if have no experiences. 
\end{enumerate}

\subsection{Interview Questions} \label{protocol}
\subsubsection{Demographic and Background Questions.}
\begin{enumerate}
    \item What is your age?
    \item How do you identify your gender?
    \item How long have you been serving as a first responder?
    \item What is your specific occupation and title?
    \item What is your duty?
    \item What are the common tasks you need to perform to fulfill your duty? Please briefly describe these tasks
    \item Could you please describe the most challenging task that you were involved in?
        \begin{itemize}
            \item Why do you think it’s challenging?
        \end{itemize}
    \item What technologies have you used to support your work? 
        \begin{itemize}
            \item How effective is the technology?
            \item What are the drawbacks of the technology?
            \item How do you want to improve it?
        \end{itemize}
    \item Have you ever used an AR glass before?
        \begin{itemize}
            \item If yes:
                \begin{itemize}
                    \item What device have you used?
                    \item What application did you use?
                    \item When did you use it?
                    \item For what purpose did you use them?
                    \item What did you like about it? Why?
                    \item What did you dislike about it? Why?
                \end{itemize}
            \item If not, what do you know about AR glasses?
        \end{itemize}
\end{enumerate}

\subsubsection{HoloLens 2 Hardware Questions.}
Participants were given a brief introduction of the HoloLens 2 hardware components. We then handed the HoloLens 2 to participants and led them wear it for three minutes. While wearing the HoloLens 2, participants were asked to imagine themselves in an emergency situation performing in-the-field tasks. They were encouraged to walk around and simulate actions common to their in-the-field tasks to understand how the glasses would function in a real-life setting while sharing their thoughts and feelings aloud with us. After three minutes, participants were asked the following questions: 
\begin{enumerate}
    \item How comfortable do you feel when wearing AR glasses? Please describe your experiences. 
        \begin{itemize}
            \item Do you feel it’s heavy? 
            \item How does wearing AR glasses affect your head or body movement?
            \item Do the AR glasses block your vision? If so, to what extent?
        \end{itemize}
    \item What do you think of the safety of wearing it while performing in-the-field tasks?
        \begin{itemize}
            \item Why? 
        \end{itemize}
    \item Do you have any concerns about wearing it during in-the-field tasks?
        \begin{itemize}
            \item What are your concerns? 
            \item Why do you have these concerns?
            \item Do you have any suggestions on improving the form factor of the glasses?
        \end{itemize}
\end{enumerate}

\subsubsection{HoloLens 2 Interaction Techniques Questions.} Participants were shown how to use the HoloLens 2 through three supported modalities included in the Tip application. We then asked the follow-up questions: 
\begin{enumerate}
    \item Do you think interaction techniques are easy to use? 
        \begin{itemize}
            \item If yes, why? 
            \item If not, 
                \begin{itemize}
                    \item What do you find difficult to use?
                    \item How would you like to improve them? 
                \end{itemize}
        \end{itemize}
    \item After familiarizing with the Hololens 2 a bit, would you like to use it during the in-the-field tasks? 
        \begin{itemize}
            \item If yes,
                \begin{itemize}
                    \item Why?
                    \item How do you think it could be useful?
                    \item In what kind of situations could it be useful? 
                \end{itemize}
            \item If not, 
                \begin{itemize}
                    \item Why not?
                    \item What are some concerns you have?
                \end{itemize}
        \end{itemize}
\end{enumerate}

\subsubsection{Brainstorming Session with Four AR Cues.} Participants were presented each type of design cues and asked the follow-up questions: 

\textbf{Overview cues:}
\begin{enumerate}
    \item Do you think the design of these maps can be helpful to you?
        \begin{itemize}
            \item If yes, 
                \begin{itemize}
                    \item How could they be useful?
                    \item In what usage scenarios do you think such maps could be useful?
                \end{itemize}
        \end{itemize}
        \begin{itemize}
            \item Do you have a preference for either 2D or 3D maps?, 
                \begin{itemize}
                    \item If so, which one works best for you? Why? 
                \end{itemize}
        \end{itemize}
    \item Are there any standard methods used by FR to present map information? 
        \begin{itemize}
            \item If yes, what are they?
            \item If not, do you think we should follow any standards? What are they? 
        \end{itemize}
    \item How do you think we can design the map to support wayfinding?
        \begin{itemize}
            \item What kind of information should the map contain? 
            \item Where do you think the map should be placed?
            \item How do you determine the optimal size and visual design of maps?
            \item How do you want to trigger the map? 
            \item Will different usage scenarios affect your answers in the design of the map? 
                \begin{itemize}
                    \item If yes, how?
                    \item If not, why?
                \end{itemize}
        \end{itemize}
    \item What are the drawbacks of these types of visual cues?
    \item How do you want to improve the map design? 
\end{enumerate}

\textbf{Highlighting cues:}
\begin{enumerate}
    \item Do you think such outlines can be useful to you?
        \begin{itemize}
            \item If yes, how could it be useful? 
            \item In what usage scenarios or tasks do you think such visual cues could be helpful? 
            \item What type of information or objects are important to highlight using such visual cues? 
        \end{itemize}
    \item Are there any standard visual cues used by FR to highlight objects in the environment?
        \begin{itemize}
            \item If yes, what are they?
            \item If not, do you think we should follow any standards? What are they? 
        \end{itemize}
    \item How do you think we can design visual cues to highlight objects? 
        \begin{itemize}
            \item Which part of an object should be highlighted? Why?
            \item What forms of design do you prefer to highlight the info/objects? Why? 
            \item How do you determine the optimal size and color of highlighting contours? 
            \item How do you want to trigger this type of visual cues?
            \item Will different objects or usage scenarios affect your answers in the design of highlighting contours? 
                \begin{itemize}
                    \item If yes, how?
                    \item If not, why?
                \end{itemize}
        \end{itemize}
    \item What are the issues with this type of visual cues?
    \item How do you want to improve such visual cues? 
\end{enumerate}

\textbf{Directional cues:}
\begin{enumerate}
    \item Do you think such a visual cue can be useful to you during wayfinding?
        \begin{itemize}
            \item If yes, how could it be useful? 
            \item What types of information should be signaled by wayfinding signs? 
        \end{itemize}
    \item Are there any standard wayfinding signs used by FR?
        \begin{itemize}
            \item If yes, what are they?
            \item If not, do you think we should follow any standards? What are they?
        \end{itemize}
    \item How do you think we can design wayfinding cues?
        \begin{itemize}
            \item What kinds of information should be involved in a wayfinding cue?
            \item Where do you think the wayfinding cues should be placed?
            \item What form of design do you prefer for wayfinding cues?
            \item How do you determine the optimal size and color of wayfinding signs?
            \item How do you want to trigger the wayfinding signs?
            \item Will different usage scenarios affect your answers in the design of wayfinding signs?
                \begin{itemize}
                    \item If yes, how?
                    \item If not, why?
                \end{itemize}
        \end{itemize}
    \item What are the drawbacks of this type of visual cues?
    \item How do you want to improve the wayfinding signs?
\end{enumerate}

\textbf{Labeling cues:}
\begin{enumerate}
    \item Do you think such symbolic representations of objects can be useful to you?
        \begin{itemize}
            \item If yes, in what usage scenarios or tasks do you think such visual cues could be helpful?
            \item What type of information or objects are suitable to be labeled by icons/symbols?
            \item What icons do you think we should use to label such information?
        \end{itemize}
    \item Are there any standard icons/symbols used by FR to indicate different objects in the environment?
        \begin{itemize}
            \item If yes, what are they?
            \item If not, do you think we should follow any standards? What are they?
        \end{itemize}
    \item Now let’s talk about the specific design of an icon.
        \begin{itemize}
            \item What kinds of information should the icons/symbols include?
            \item If we want to use the icon to label an object (e.g., an entrance, an injured person), where do you think the symbols should be placed?
            \item How do you determine the optimal size and color of icons/symbols?
            \item How do you want to trigger this type of visual cues?
            \item Will different target objects or usage scenarios affect your preference for  the design of icons/symbols?
                \begin{itemize}
                    \item If yes, how? 
                    \item If not, why?
                \end{itemize}
        \end{itemize}
    \item What are the issues with this type of visual cues?
    \item How do you want to improve it?
\end{enumerate}

\subsubsection{Final Interview.}
\begin{enumerate}
    \item In general, do you think AR glasses would help your tasks as a first responder, especially in the in-the-field tasks? 
        \begin{itemize}
            \item What are the tasks that you think AR glasses can support? How?
            \item What are the benefits of using AR glasses?
        \end{itemize}
    \item Would you be willing to use them in the future?
        \begin{itemize}
            \item Why or why not?
            \item What are your concerns about using them via AR glasses?
        \end{itemize}
    \item What are the visual cues that you would prioritize and always present them in your view, if there are any?
        \begin{itemize}
            \item Why?
            \item How would you like them to present constantly in your view?
        \end{itemize}
    \item What are the visual cues that should be switched on and off?
        \begin{itemize}
            \item Why are they optional? In what scenarios?
            \item How would you like them to be triggered?
        \end{itemize}
    \item Can you think of any other visual cues that could be used to support tasks of first responders via AR glasses?
        \begin{itemize}
            \item If yes, 
                \begin{itemize}
                    \item What are they? 
                    \item For what tasks?
                    \item Why are they important?
                    \item How could they be useful? 
                \end{itemize}
        \end{itemize}
    \item Apart from the tasks we’ve discussed, what are some other FR tasks that AR glasses could assist with?
        \begin{itemize}
            \item Please elaborate the tasks with examples
            \item How could AR glasses be helpful in these tasks?
        \end{itemize}
    \item Apart from the visual cues studied in this study, what are some other cues that might be useful for you during the tasks? [prob: go over audio feedback, haptic feedback, hand gestures for the following questions]
        \begin{itemize}
            \item Do you think XXX will be helpful? 
                \begin{itemize}
                    \item If so,
                        \begin{itemize}
                            \item How could it be helpful? 
                            \item Why do you think XXX is useful?
                            \item In what usage scenarios do you think it could be helpful?
                            \item For what kinds of tasks?
                            \item Can you estimate some drawbacks of XXX?
                        \end{itemize}
                    \item If not, why? 
                \end{itemize}
        \end{itemize}
    \item How do you want to interact with and control AR glasses? [prob: hand gestures? Eye tracking? Speech commands?]
        \begin{itemize}
            \item Why do you want to interact with it via this way?
            \item In what tasks?
        \end{itemize}
    \item What are the improvements (or ideal designs) you would like to see in future AR glasses to support first responders?
    \item Is there anything else that you’d like to share apart from the things that we have already discussed?
\end{enumerate}

\subsection{Theme Table}
\begin{table*}[h!]
\centering
\footnotesize 
\caption{Themes and Codebook.}
\begin{tabular}{p{2.5cm}p{3.5cm}p{7cm}}
\toprule
\textbf{Themes} & \textbf{Sub-themes} & \textbf{Codes} \\
\midrule
Hardware Form Factor & Limited field-of-view & bifocal view; periphery blocks vision; tunnel vision; head need to be in the right position; full streamlined glasses \\ 
\cmidrule{2-3}
& Extra stability for role-specific needs & comfortable weight; no limit to movement; gender-specifc needs; fatigue/headache for long term wearing; unstable with big body movement; add chin strap \\
\cmidrule{2-3}
& Integration to existing equipment  & incompatible with helmet; difficult to wear in small spaces; social acceptability; privacy concerns; refine form factor; OSHA standards for hardware \\
\midrule

Overview Cues & 2D for urgency vs. 3D maps for training & prefer 3D over 3D; accurate 3d maps help see things at different angles; no standard to present map information; 3D helping training simulation; trade-off between information and simplicity \\
\cmidrule{2-3}
& Multi-layered map with structural base and on-demand layers & add switchable info layers; be easily (de)activated based on needs; base layer should always be the same; hazards layers; see teammate's positions; navigation support; no standard navigation signs; help spatial orientation; allow customization of size, color, position;  \\
\midrule

Directional Cues & Back-tracing FF in visually challenging conditions & arrows for back tracing; dark navigation; dense smoke\\
\cmidrule{2-3}
& Subtle cues for driving & emergency vehicle driving; locate suspects; trade-off between visibility and intrusiveness; augmentation immersed in the environment; attention switching; enhanced distance perception; avoid blocking real-world view \\
\cmidrule{2-3}
& Real-time environment marking for collaboration & leverage prior knowledge; sync info among collaborating parties; \\
\cmidrule{2-3}
& AR compass for outdoor navigation & outdoor navigation; need compass outside of the building \\
\midrule

Highlighting Cues & Outlines vs. shades & shade indicating hazard (not go through); standard design of outlines; avoid information overload; prefer shades to highlight object; blurry shades\\
\cmidrule{2-3}
& Outline informing object affordance & door swing direction; inform actions; security like a burglar wire or panic hardware; incorporate thermal imaging lens \\
\cmidrule{2-3}
& Shades indicating object properties & want to see the entirety of the door; know door size; inform the exact objects from shape \\
\midrule

Labeling Cues & Label small or hidden hazards &label hazard prevention and control resource; save searching time; detect warm bodies; visualize hidden hazards; identify weapon on people\\
\cmidrule{2-3}
& Follow standard icons in the field &adopt current standard symbols or icons; (e.g.,) KNOX Box with KB, FDC as a fire department connection; cues at eye level; follow CADD standards\\
\cmidrule{2-3}
& Label human subjects on scene & detect people's identity on the scene; highlight target among crowded vision \\
\midrule

Properties Preferences & Colors; & adaptive color based on real-world background; virtual triaging; color representation; prefer neutral color\\
\cmidrule{2-3}
& Thickness & use thickness to denote urgency level; thicker for more urgent \\
\cmidrule{2-3}
& Opacity & opacity levels; shallower when further away; rapidly changing priority\\
\cmidrule{2-3}
& Placement & place arrows on the ground; crawling; smoke-filled environment \\
\midrule

Multi-modal Interaction & Speech commands; & hands-free; voice command be unique and short; avoid interrupting teammates; challenging to remember voice commands \\
\cmidrule{2-3}
& Gaze interaction & dislike gaze interaction; cause distraction from the task focusing on; like eye tracking; avoid extra head movement; hands-free \\
\cmidrule{2-3}
& Tangible controls & tactile confirmation; big button \\
\cmidrule{2-3}
& In-the-air gestures & pull up maps; hard to do with gloves; occupied hands \\
\cmidrule{2-3}
& Head vibration & add haptic feedback; stronger vibration when approaching hazards\\

\bottomrule
\end{tabular}
\label{tab:themes}
\end{table*}

\end{document}